\documentclass[fleqn,usenatbib]{mnras}

\usepackage{newtxtext,newtxmath}
\usepackage{threeparttable}

\usepackage[T1]{fontenc}

\DeclareRobustCommand{\VAN}[3]{#2}
\let\VANthebibliography\thebibliography
\def\thebibliography{\DeclareRobustCommand{\VAN}[3]{##3}\VANthebibliography}

\usepackage{graphicx}	
\usepackage{amsmath}	
\usepackage{gensymb}    

\title[A nuclear disc at Cosmic Noon]{A nuclear disc at Cosmic Noon: evidence of early bar-driven galaxy evolution}

\author[Z. A. Le Conte et al.]{Zoe A. Le Conte,$^{1}$\thanks{E-mail: zoe.a.le-conte@durham.ac.uk}
Dimitri A. Gadotti,$^{1}$
Thomas Harvey,$^{2}$
Leonardo Ferreira,$^{3}$
Christopher J. Conselice,$^{2}$
\newauthor
Taehyun Kim,$^{4}$
Camila de Sá-Freitas,$^{5}$
Francesca Fragkoudi,$^{6}$
Justus Neumann,$^{7}$
and
E. Athanassoula$^{8}$
\\
$^{1}$Centre for Extragalactic Astronomy, Department of Physics, Durham University, South Road, Durham DH1 3LE, UK\\
$^{2}$Jodrell Bank Centre for Astrophysics, University of Manchester, Oxford Road, Manchester M13 9PL, UK\\
$^{3}$Instituto de Matemática Estatística e Física, Universidade Federal do Rio Grande, Rio Grande, RS, Brazil\\
$^{4}$Department of Astronomy and Atmospheric Sciences, Kyungpook National University, Daegu, 41566, Republic of Korea\\
$^{5}$European Southern Observatory, Alonso de C\'ordova 3107, Vitacura, Regi\'on Metropolitana, Chile\\
$^{6}$Institute for Computational Cosmology, Department of Physics, Durham University, South Road, Durham DH1 3LE, UK\\
$^{7}$Max-Planck-Institut f\"{u}r Astronomie, K\"{o}nigstuhl 17, D-69117 Heidelberg, Germany\\
$^{8}$Aix Marseille Univ, CNRS, CNES, LAM, Marseille, France\\
}

\date{Accepted 2026 June 10. Received 2026 June 9; in original form 2026 January 23}

\pubyear{\the\year{}}

\begin{document}
\label{firstpage}
\pagerange{\pageref{firstpage}--\pageref{lastpage}}
\maketitle

\begin{abstract}
Recent studies have revealed that bars can form as early as a few billion years after the Big Bang, already displaying characteristics similar to those of evolved bars in the Local Universe. Bars redistribute angular momentum throughout the galaxy, regulating star formation, AGN activity, and the formation of new stellar structures such as nuclear discs. However, the effects of bar-driven evolution on young galaxies are not yet known, as no evidence of bar-built stellar structures has ever been found beyond $z = 1$, until now. In this work, we present evidence for a bar-built, star-forming nuclear disc already present at redshift $z = 1.5$. This is the first evidence of a bar-built stellar structure at Cosmic Noon. We find that this nuclear disc is actively forming stars and is of similar size to some nuclear discs in nearby galaxies. This evidence solidifies the now emerging picture in which bars are fundamental not only in the late evolution of galaxies, but also in their early evolutionary stages. It changes the current paradigm by urging a revision of our picture of galaxy evolution beyond redshift one to include new considerations of the role of bars as early as a few billion years after the Big Bang.
\end{abstract}




\section{Introduction}

Stellar bars are ubiquitous elongated structures in nearby disc galaxies, and are the key drivers of internal secular galaxy evolution; hence, the formation of a bar is associated with the instance of ``disc settling'' \citep[e.g.,][and references therein]{Coelho_2011,Keyes_2019,Gadotti_2020,Silva-Lima_2022,Garland_2024}. Studies of barred galaxies at redshifts $z > 1$ have become prevalent in the past few years, thanks to the improved sensitivity and infrared coverage of the James Webb Space Telescope (JWST). These studies have revealed that bars can form as early as a few billion years after the Big Bang, with the fraction of disc galaxies hosting a bar at Cosmic Noon ranging from $\sim 10$ to 20 per cent \citep[e.g.,][]{LeConte_2024,LeConte_2025,Guo_2023,Guo_2025,Géron_2025,Salcedo_2025,euclid2_2025}. Remarkably, studies have revealed that these bars at high redshifts have characteristics similar to those of nearby barred galaxies, indicating that these structures are established, highly evolved, and ``mature''. In a sample of $\sim70$ barred galaxies at $z \geq 1$, \citet{LeConte_2025} measured via ellipse fits that these bars are as long as those in the local Universe, and that bars grow in tandem with their discs. Some studies have suggested that young bars have a characteristic exponential surface brightness profile \citep{Kim_2015}, but in a sample of nine massive barred galaxies at $z \approx 1.5$, \citet{Kalita_2026} identified only two bars to have this feature, whereas over half of the sample were shown to be highly evolved systems with flat bar profiles. \citet{Pastras_2025} found evidence of rapid gas inflow along the bar of a star-forming galaxy at $z \approx 1.5$, and \citet{Salcedo_2025} reported that barred galaxies at Cosmic Noon are highly rotationally supported with values of $V/\sigma > 5$. The formation mechanism of bars across cosmic time remains observationally inconclusive; however, at $z \geq 2$, simulations find that tidal interactions may dominate \citep[e.g.,][]{Bi_2022,Lokas_2025,Zheng_2025}, and \citet{Tsukui_2024} find evidence of a distant ($z = 4.4$) barred galaxy formed via an interaction.

Nuclear discs are frequent structures in local barred galaxies and range in size and properties \citep[e.g.,][see also \cite{Schultheis_2025} for a review]{Erwin_2004,Comeron_2010}. The sizes of nuclear discs are on the order of hundreds of parsecs in radius, and are rapidly rotating structures at the centre of the barred galaxy, separate from the main galaxy disc. As such, nuclear discs have properties that contrast with those of classical bulges, and the two structures appear in observations of nearby galaxies at least as frequently \citep[e.g.,][]{Bittner_2020,Fraser_2025}. A nuclear disc is thought to be formed by the bar funnelling gas towards the centre, whereby a gas reservoir forms; thus, igniting star formation in the central kiloparsec of the galaxy \citep[e.g.,][]{Bittner_2020}. The nuclear disc size is limited by the inner Lindblad resonances (ILR) of the bar potential \citep[][]{Shlosman_1989}, and recent studies have reported on the relation that nuclear discs in the local Universe are $\sim 13 \%$ of their bar lengths \citep[][]{Gadotti_2020}. The nuclear disc is an indicator that bar-driven galaxy evolution has commenced and is effectively redistributing angular momentum. Moreover, it creates a stable environment for further gas funnelling toward the central supermassive black hole, and potentially fuelling AGN. Further gas inflow is achieved by the formation of non-axisymmetric nuclear substructures in the nuclear disc, such as a nuclear bar or spiral arms. The impact of bars and the occurrence of nuclear discs during Cosmic Noon remain unexplored due to spatial-resolution limitations \citep[see the discussion in][]{Gadotti_2026}.

In this Letter, we present detailed structural analyses of JWST images of a barred galaxy, providing evidence for a nuclear disc at a spectroscopic redshift of $z = 1.5$. This is the most distant nuclear disc to date. In the next section, we present the data and analyses performed; in \S~\ref{Sec:results}, we present the results, which are discussed in \S~\ref{Sec:discussion}. Section \ref{Sec:conclusions} summarises our main conclusions. We adopt the Planck 2020 cosmological parameters for a flat $\Lambda$CDM cosmology with H$_{0}$ = 67.36, $\Omega_{m}$ = 0.3153, and $\Omega_{\Lambda}$ = 0.6847 \citep{Planck_2020}.

\section{Data analysis}
\label{Sec2}
\subsection{Data}
For this study, we utilise public JWST observations from the Cosmic Evolution Early Release Science Survey \citep[CEERS; PI: Filkelstein, ID=1345,][]{Finkelstein_2023}. These observations included imaging from the Near Infrared Camera (NIRCam) filters (F115W, F150W, F200W, F277W, F356W, F410M, and F444W) and spectra from the Near Infrared Spectrograph (NIRSpec) with medium resolution, $R \sim 1,000$, and gratings (G140M, G235M, and G395M). The NIRCam exposures we retrieved from the \texttt{Mikulski Archive for Space Telescopes (MAST)} and reprocessed with the official JWST pipeline (v1.8.2, CRDS v1084), following the procedures of \citet{Ferreira_2022} and \citet{Adams_2023} but incorporating several refinements which are summarised in \citet{LeConte_2025}. To ensure sub-pixel alignment across the dataset, each filter is reprojected onto the F444W grid at $0.03^{\prime\prime}$ pixel$^{-1}$. A comprehensive description of the workflow and its validation is given in \citet{Adams_2024, Conselice_2024, Harvey_2024}. We thus produce $30 \ \rm mas$ 128x128 pixel$^2$ cutouts from the NIRCam data.

The reduced spectra for the galaxy were obtained from the DAWN JWST Archive\footnote{https://dawn-cph.github.io/dja/} (DJA). A consistent redshift of $z=1.461$ was identified across all 3 spectra (G140M, G235M, G395M) using the \texttt{msaexp} redshift fitting tool (which the DJA uses). The spectra cover the nuclear bar, but analysing the spectra is beyond the scope of this paper, since spectra of the main bar or disc are not available. In this study, we use the NIRCam empirical Point Spread Function (PSF) Full Width Half Maximum (FWHM)\footnote{JWST user documentation: https://jwst-docs.stsci.edu/jwst-near-infrared-camera/nircam-performance/nircam-point-spread-functions}. The F115W PSF FWHM is $0.040^{\prime\prime}$; corresponding to a linear resolution of 0.347 kpc for $z = 1.461$. The PSF FWHM degrades for longer-wavelength filters; hence, the F444W PSF FWHM is $0.145^{\prime\prime}$, corresponding to a linear resolution of 1.259 kpc. Our analysis is primarily conducted on the F150W and F200W filters, with PSF FWHMs of $0.050^{\prime\prime}$ and $0.066^{\prime\prime}$, corresponding to linear resolutions of 0.434 and 0.573 kpc, respectively.

\subsection{Unsharp masking}
Bright components at the centre of galaxies can make it difficult to identify finer substructures, such as the nuclear disc. The change in light between the nuclear disc and the stellar bar or main disc can be enhanced by the effective technique of unsharp masking \citep[e.g.,][]{Malin_1977}, which removes light associated with large structures in the galaxy. We generate unsharp masked images by first convolving the NIRCam image with a circular Gaussian kernel with $\sigma = 4.4$ pixels, corresponding to $0.132^{\prime\prime}$. The kernel size was chosen to be greater than $2\times$FWHM and the radius of a possible nuclear disc. The original NIRCam image is then divided by the convolved image to reveal small structures in the galaxy.

\subsection{Photometric decompositions}
\label{Sec: decomp}
A powerful technique for deriving the structural properties of a galaxy is to perform a two-dimensional photometric decomposition, which we use here with \textsc{IMFIT} v1.8 \citep{Erwin_2015}. We prepared the NIRCam images for \textsc{IMFIT} employing \textsc{SExtractor} \citep[][]{Bertin_1996} to create masks of neighbouring sources. Additionally, we convert the pixel units of the NIRCam images from MJy/sr to DN/s by dividing the image by the \texttt{FLUXCONV} constant from the image header. We edit the \textsc{IMFIT} configuration file to have the corresponding gain, readout noise, subtracted background, and exposure time for each NIRCam filter. The galaxy centre is defined as the brightest pixel and is determined using ellipse-fitting techniques (described below). Furthermore, we provide \textsc{IMFIT} with a central pixel limit of $\pm 2$ pixels from this specified value. The fitted structural components are centred on the galaxy centre. We fed \textsc{IMFIT} with simulated NIRCam PSFs generated by the \textsc{STPSF} package \citep{Perrin_2014}. We selected a region of $\pm 20$ pixels centred on the galaxy to convolve with a fourfold-oversampled PSF, and we used the Differential Evolution (DE) algorithm \citep[][]{Stron_1997} to minimise the fit statistic and find the best-fit model. The algorithm does not require an initial guess for parameter values; rather, it requires upper and lower limits, which makes it less likely to be trapped in local minima \citep[for a discussion on the robustness of the DE algorithm, see][]{Gadotti_2026}.

We fit all NIRCam images with three components. The radial light/mass profile of the disc is described by an exponential; a generalised S\'ersic function describes the bar; a S\'ersic function describes the central component. Collectively, we fit 18 free parameters. As a result, we have a component model defined by concentric and homologous ellipses of position angle PA and ellipticity $\varepsilon$, defined by 
\begin{equation}\label{ell}
\varepsilon = 1 - \frac{b}{a} ,
\end{equation}
where $b$ is the length of the semi-minor axis and $a$ is the length of the semi-major axis of the ellipse. The intensity for a given ellipse in the central component is defined as:
\begin{equation}\label{eq1}
    I(r) = I_{e} \exp \left( -b_{n} \left[ \left( \frac{r}{r_e} \right) ^{1/n}-1 \right] \right),
\end{equation}
where $I_e$ is the intensity at the effective radius $r_e$, $n$ is the S\'ersic index and $b_{n}$ is a constant that is a function of $n$. We use these models to obtain the structural parameters of the different galactic components, such as characteristic sizes and S\'ersic index, as well as luminosity/mass fractions. In addition, we subtract the combined models from the original NIRCam image to create a residual image that enhances unmodelled stellar structures.

\subsection{Isophotal analysis}
Structures in galaxies can also be traced with contours of equal intensity, i.e., isophotes. The structures deduced from isophotal contours can be parameterised with ellipse fitting. For example, an indication of a barred nuclear disc in a barred galaxy is a double-peaked radial profile of isophotal ellipticity \citep[see][]{Erwin_2004}. 

We use \texttt{photutils.isophote} from Python's astropy package \citep[e.g.,][]{Bradley_2022} to fit elliptical isophotes to the NIRCam images. Firstly, we run this module with the galaxy centre coordinates as free parameters, and iterate through the initial parameters to obtain the best fit. The module assesses a $10 \times 10$ window centred around a specified point for each isophote to get a central coordinate. Then, we select isophotes with radii of $10 - 40$\% of the whole galaxy, and determine the galaxy centre as the average of their central coordinates. Finally, the isophotal ellipse fitting is repeated, however, with a fixed centre.

We obtain the parameters of the fitted ellipses at increasing radii, and we are most interested in the radial profiles of the isophotal ellipticity. We thus focus on identifying the semi-major axis of any major, isolated ellipticity peaks.

\subsection{SED fitting}
\label{Sec: SED}
We perform resolved SED fitting using the Bayesian SED fitting tool \textsc{bagpipes} \citep[e.g.,][]{carnall2018} via the \textsc{expanse} package \citep[e.g.,][]{harvey2025} in order to produce resolved measurements of the stellar mass and star formation rate surface densities in the central component, bar, and disc of the galaxy. Our SED fitting model is constructed as follows: we fix the redshift to the known spectroscopic redshift, with the flexible non-parametric `continuity’ star formation history (SFH) of \citet{leja2019}. To model dust attenuation, we use a single-component dust law \citep{Calzetti_2000}, with a uniform prior of $0 \leq \rm A_V \leq 5$. We utilise a BPASS \citep{stanway2018} SPS model, with a \citet{kroupa2001} initial mass function (IMF), allowing for nebular emission from star-forming regions using CLOUDY post-processing with a uniform prior on the logarithm of the ionisation parameter $\log U$ between -4 and -1. We allow a broad log-uniform prior on stellar metallicity, between $10^{-3}$ and $2.5 Z_\odot$. 

Our resolved SED fitting process broadly follows \citet{harvey2025}; in short, we convolve all NIRCam cutouts to the measured PSF for the F444W filter using an empirical PSF model, then perform Voronoi binning on the F277W cutout using \textsc{vorbin} \citep{cappellari2012} to ensure a per-bin SNR $> 10$ in every fitted region. This SNR is sufficient for simple measurements of the D4000, but in the central regions, which are the focus of this study, it is largely exceeded. We mask a small number of bins very close to the detector edge in the long wavelength filters, which are noise-dominated. We fit each bin independently using \textsc{bagpipes} as described above using MultiNest nested sampling \citep[e.g.,][]{feroz2009}. We probe the strength of the 4000\AA\ break (D4000) by also using HST ACS/WFC F606W ($\sim 0.6\mu$m) and F814W ($\sim 0.8\mu$m). We thus produce maps of star formation rate (SFR) density, stellar mass density, and the D4000, corresponding to the 50th percentile of the posterior distribution for each bin and parameter.

\section{Results}
\label{Sec:results}
Armed with the analysis tools described above, we report here the discovery of the first nuclear disc observed beyond $z = 1$, with an age of only 4.5 billion years after the Big Bang. The $\log_{10}(M_\star) = 10.75\pm 0.05$ M$_\odot$ galaxy was categorised as a strongly barred galaxy in our study of the bar fraction \citep[e.g.,][]{LeConte_2025}, which used visual classification and ellipse fits to identify a high-redshift sample of barred galaxies and measure their properties. This galaxy already stood out in that analysis, since we measured a long bar with projected length (semi-major axis) $L_{bar} = 5.33 \pm 0.08$ kpc in a near face-on spiral disc. However, we note that given the low inclination of the galaxy, the true, deprojected length is not substantially different, although it will naturally be slightly larger.

\begin{figure*}
\centering
\includegraphics[width=0.99\textwidth]{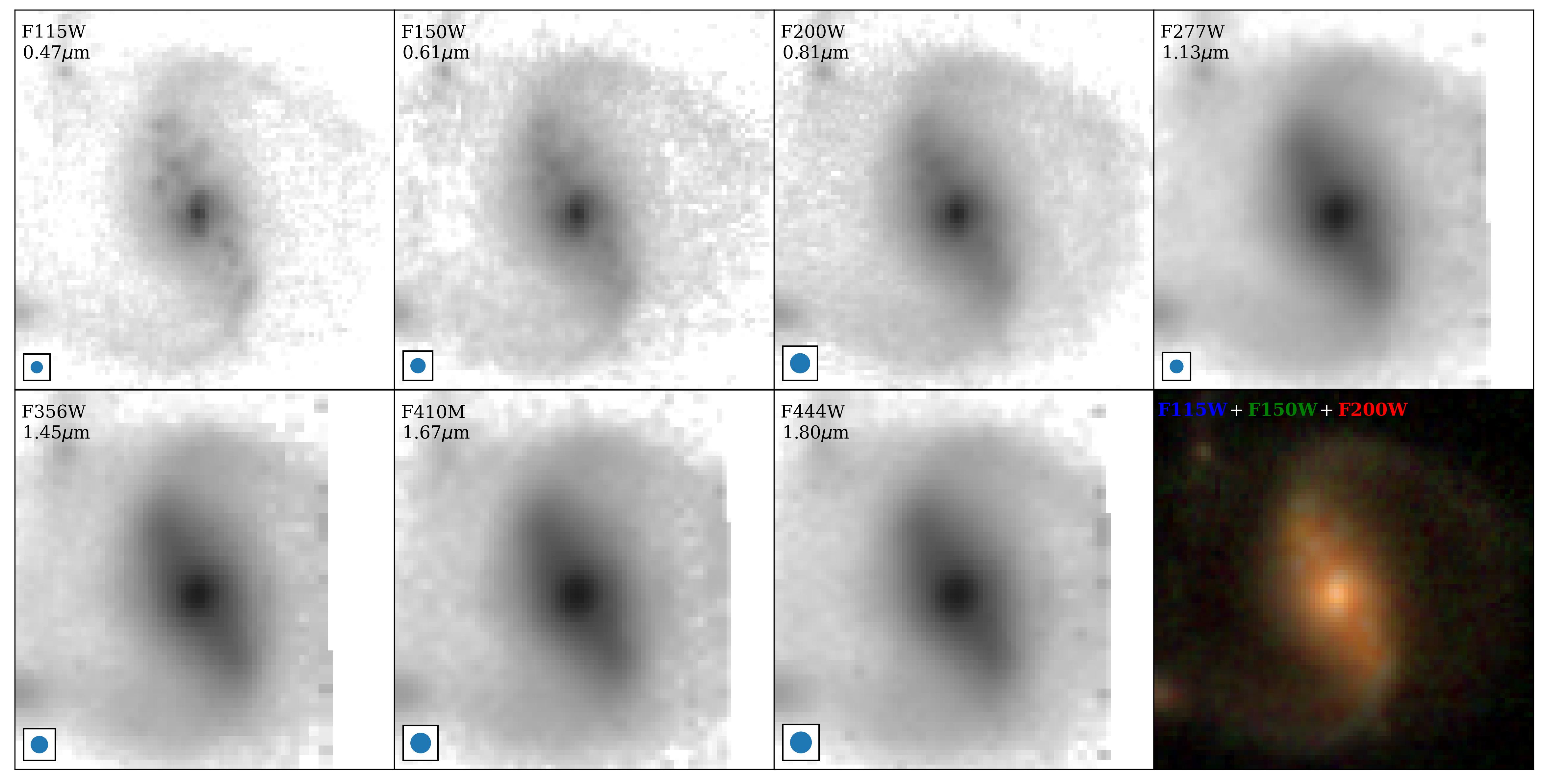}
\caption{The galaxy images from seven NIRCam filters, annotated in the top-left corner of each image with the filter name and rest-frame wavelength for a redshift of $z = 1.461$. A circle depicting 2×FWHM of the PSF is shown in the lower-left corner of each image. The lower-right panel is an RGB image obtained from the filters F115W, F150W and F200W.}
\label{fig1}
\end{figure*}

Figure \ref{fig1} shows the galaxy in seven NIRCam filters, where a central structure is already discernible; in particular, an elongated structure is seen in the bluest filters (which have better spatial resolution). The NIRCam filter wavelength range spans from $1.15 \mu$m to $4.44 \mu$m, corresponding to a rest-frame wavelength of $0.47 \mu$m to $1.80 \mu$m for this galaxy at $z = 1.461$. The RGB image, constructed using images from the shorter-wavelength channel, reveals that the ends of the elongated central stellar structure are bluer, indicating younger stellar populations and enhanced star formation.

Using three of the techniques described in detail in \S~ \ref{Sec2}, encompassing structure-enhancement and model-dependent approaches, we reveal the nature of the nuclear structure. These techniques are: unsharp masking, which enhances high spatial frequency substructures; 2D photometric decomposition, which involves the fitting of a three-component model composed of a disc, bar, and a central component representing the main stellar structures of the galaxy; and isophotal ellipse fitting to the isoluminance contours of the image, allowing one to obtain a radial ellipticity profile of the galaxy and its inner components. The photometric decomposition can also be used to produce residual images, often revealing substructures.

\begin{figure*}
\centering
\includegraphics[width=0.99\textwidth]{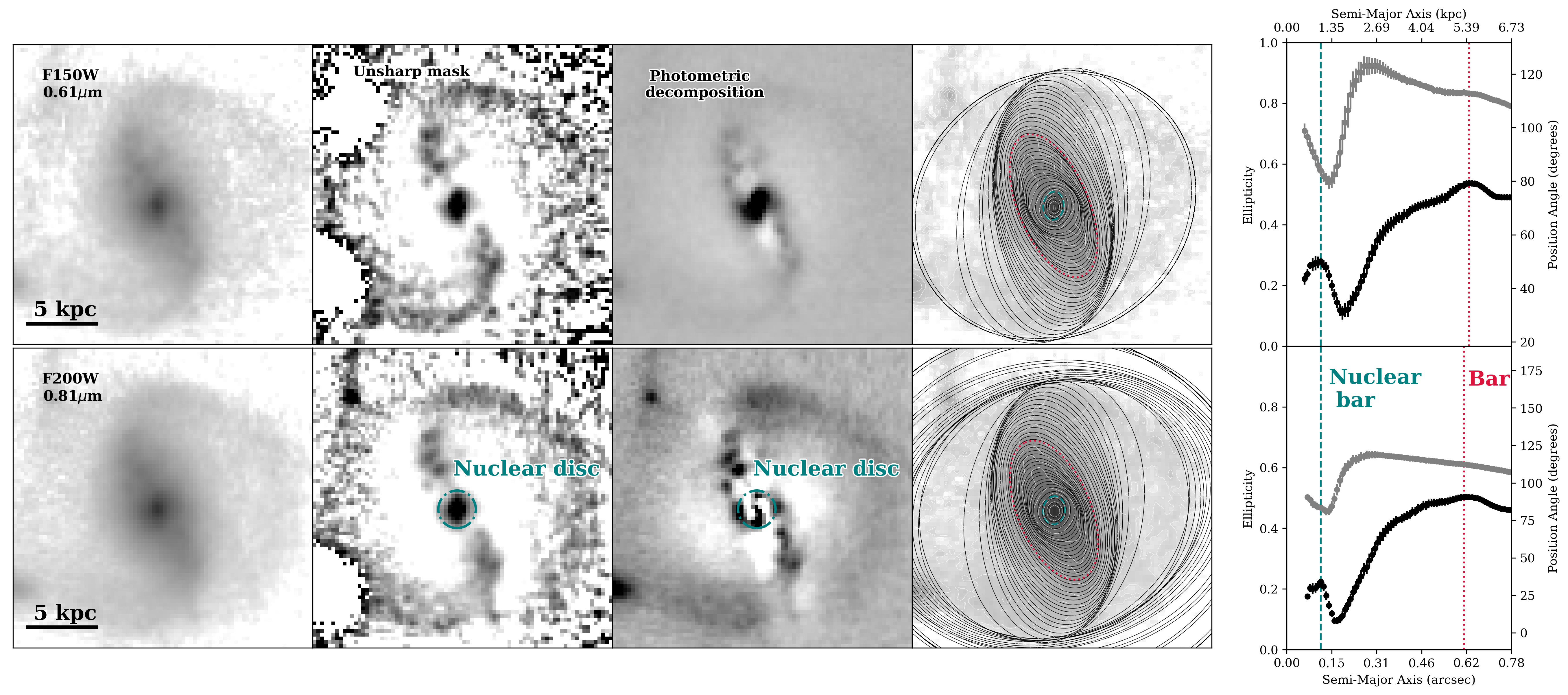}
\caption{Image analysis of the galaxy in the F150W (top row) and F200W (bottom row) NIRCam filters. From left to right: NIRCam image; unsharp masked image; IMFIT residual image for a multi-component fit; isophotal ellipse fitting of the NIRCam image; ellipticity (black) and position angle (grey) radial profile from ellipse fitting, showing the peak in ellipticity of the nuclear bar in the nuclear disc as a dashed line and of the main bar as a dotted line. The nuclear disc size, measured by visual inspection of the images, is shown as a dot-dashed circle in the unsharp-masked and residual images.}
\label{fig2}
\end{figure*}

The results from the analyses of the NIRCam filters F150W and F200W are shown in Figure \ref{fig2}. These two filters were chosen for their optimal balance between high spatial resolution and not being too `blue', where the main bar becomes too clumpy and a fit cannot be obtained. An elongated structure exhibiting a pair of symmetric `wings' resembling spiral arms $\approx 1.3$ kpc in radius, is enhanced in the unsharp-marked images. These images reveal a distinctive light deficit between the bar and this central stellar structure. The light deficit is important because it indicates a physical separation between the central stellar structure and the bar, and thus is not simply a continuation of the bar towards the centre.

The stellar substructure is further highlighted in the residuals of the multi-component fit to the image. For the NIRCam filters F150W and F200W, we subtract the optimised model from the original image, revealing a spiral structure. Again, the nuclear spiral structure consists of two winding arms that extend to $\approx 1.3$ kpc. In F200W, the 2D decompositions indicate a central component with an effective radius, $R_{e} \approx 0.5$ kpc, and with a S\'ersic index, $n = 0.8$. In F150W, we measure an effective radius, $R_{e} \approx 0.4$ kpc, and a S\'ersic index, $n = 0.5$, although note that slight differences between filters may reflect the different dominant stellar populations in each filter. The near-exponential radial profile of the central component strongly suggests that it is indeed a nuclear disc hosting nuclear spiral arms, much like nearby counterparts such as NGC 1097 and NGC 4314 \citep[see][]{Erwin_2003,Gadotti_2019}. 

The uncertainties in parameters derived from photometric decomposition are non-trivial; however, studies have investigated the extent of these errors. Firstly, the decision on which model to use and how many structural components to include is guided by DE fits, which are statistically justified by the \citet{Akaike_1974} information criterion (AIC). \citet{Kim_2014} discussed the variation in structural properties when changing the sky background by $1\sigma$, and found that an increased (decreased) sky background increased (decreased) the Bar/T and ND/T by $\sim 8$\% (see their Sec. 3.2). A variation in the sky value affects the bar and central component effective radius, $R_e$, by less than 3\%. Furthermore, the author derived \textit{statistical} errors\footnote{\citet{Kim_2014} derives \textit{statistical} errors from \textsc{budda} \citep{deSouza_2004,Gadotti_2008} which gives robust $1\sigma$ uncertainties on each structural parameter. Whereby, the programme finds the global $\chi^2$ minimum, and in keeping all other parameters fixed, iterates until the $\chi^2$ values equate to a variation in $1\sigma$.} of $< 10$\% for the nuclear disc effective radius, 17\% for the bar effective radius, with ellipticities ranging from 5\% to 15\% for these two components, 13\% for the central component S\'ersic index, and 24\% for the bar S\'ersic index. Thus, the $1\sigma$ error corresponds to $\sigma = 0.11$ for $n = 0.843$ and $\sigma = 52.5$ for $R_e = 525$. The recent work of \citet{Gadotti_2025} conducts nuclear disc structural analysis on the TIMER sample \citep[][]{Gadotti_2019} and reports that nuclear disc S\'ersic indexes range from $\approx 0.5$ to $\approx 2$, peaking at $\approx 1.5$. Additionally, the study found the nuclear disc size spans from 200 pc to 900 pc, peaking at 500 pc. We find our nuclear disc at Cosmic Noon to be similar to, if not larger than, those observed in nearby galaxies.

The elongated central structure from which the nuclear spiral arms emerge, readily seen in the NIRCam images taken with the bluest filters, suggests that the nuclear disc also hosts a nuclear bar. An indication of a barred nuclear disc within a nearby barred galaxy is shown as a double peak in the radial ellipticity profile derived from ellipse fits, since the two bars produce independent ellipticity peaks. Indeed, we do observe this behaviour in the NIRCam F150W and F200W filters. The peak of the nuclear bar ellipticity, $\varepsilon_{\rm F150W} = 0.28 \pm 0.02$ and $\varepsilon_{\rm F200W} = 0.22 \pm 0.01$, occurs at the same radii in both filters. Thus, corresponding to the peak in ellipticity, the nuclear bar has a projected length of 1050 pc, and the radial profile is derived with a step size of 78 pc; hence, the nuclear bar size is $L_{NB} = 1050 \pm 78$. The shorter wavelength F150W filter indicates a slightly higher ellipticity for the nuclear and main bars than in the F200W filter. Such an effect has also been seen in local barred galaxies \citep[e.g.,][]{Delmestre_2024}. Lower ellipticity measurements derived via ellipse fitting in the central region could be due to PSF effects that make the isophotes rounder. Furthermore, the nuclear disc alone would help lower this parameter relative to its true value \citep[see Fig. 6 in][]{Gadotti_2008}. Hence, the true ellipticity of the nuclear bar is likely to be significantly higher. The true ellipticity of the nuclear bar could be derived from a nuclear bar model in the photometric decomposition, but this may prove unfeasible due to the low physical spatial resolution. On the other hand, we cannot rule out, based on the current analysis, that the alignment of star-forming regions could produce the elongation we interpret as a nuclear bar. The PA is not exactly constant in the NSD/NB region but shows a minor variation of about 10 degrees. It is unlikely that this is caused by the nuclear spiral arms because the isophotes corresponding to the latter are quite round, as the galaxy is close to face-on, and so their PA is ill-defined, and their ellipticity is low. The nuclear disc itself is slightly larger than the nuclear bar, as seen often in nearby galaxies. In the two approaches that we have conducted, the measurements agree that the nuclear disc size is approximately 1.3 kpc. We list the structural properties of the nuclear disc derived from our analyses along with the host galaxy properties in Table \ref{tab1}.

\begin{table}
    \centering
    \begin{threeparttable}[b]
        \begin{tabular}{@{}lll@{}}
            \hline
            Property & Value & Unit \\
            \hline
            RA & 214.937795 & $\degree$ \\
            DEC & 52.826470 & $\degree$ \\
            Redshift ($z$) & 1.461 & \\
            Stellar mass ($\rm \log_{10}(M_\star/M_\odot)$)\tnote{a} & $10.75\pm 0.05$ & \\
            Bar length ($L_{bar}$)\tnote{b} & $5.33 \pm 0.08$ & kpc \\
            Bar ellipticity ($\varepsilon_{bar}$)\tnote{b} & $0.50 \pm 0.01$ &  \\
            Nuclear bar size ($L_{\rm NB}$)\tnote{b} & $1050 \pm 78$ & pc \\
            Nuclear disc size ($R_{\rm ND}$)\tnote{c} & $1300$ & pc \\
            Nuclear disc effective radius ($ R_{e}$)\tnote{d} & $525 \pm 52.5$ & pc \\
            Nuclear disc S\'ersic index ($n$)\tnote{d} & $0.843 \pm 0.11$ &  \\
            Disc-total luminosity ratio ($D/T$)\tnote{d} & $0.50 \pm 0.08$ &  \\
            Bar-total luminosity ratio ($Bar/T$)\tnote{d} & $0.33 \pm 0.03$ & \\
            Nuclear disc-total luminosity ratio ($ND/T$)\tnote{d} & $0.17 \pm 0.01$ &  \\
            \hline
        \end{tabular}
        \caption{Main properties of CEERS-4031.}
        \label{tab1}
        \begin{tablenotes}
            \item [a] Photometrically derived from the same SED fitting model as described in \S~\ref{Sec: SED}.
            \item [b] From ellipse fits in F200W.
            \item [c] From visual inspection of the unsharp-masked images and decomposition residuals in F200W. Both measurements agree on the quoted value.
            \item [d] From the photometric decomposition in F200W.
        \end{tablenotes}
    \end{threeparttable}
\end{table}

\begin{figure}
\centering
\includegraphics[width=0.99\columnwidth]{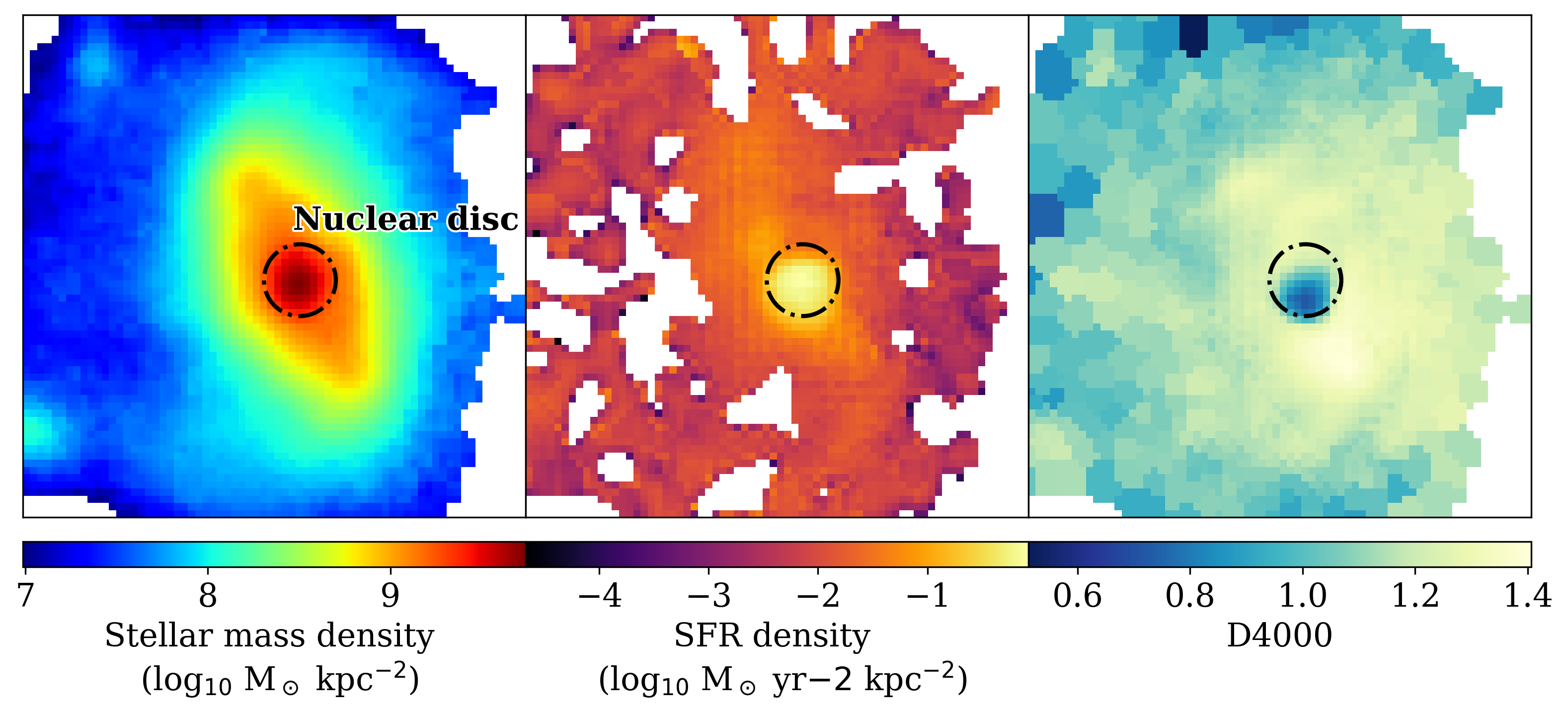}
\caption{Resolved property maps from NIRCam SED fitting. Left to right: stellar mass density, SFR density, and the strength of the 4000\AA\ break.}
\label{fig3}
\end{figure}

In local galaxies, the outer parts of nuclear discs are often regions of heightened star formation and younger stellar populations because gas is driven towards the centre by the bar \citep[see also][]{Huang_2025}. To examine whether the area we have identified as the nuclear disc exhibits this characteristic, we perform a resolved SED fitting and produce property maps, shown in Figure \ref{fig3}, observing a high SFR density within $R_{\rm ND} = 1.3$. Additionally, we find very low D4000 values within $R_{ND}$, indicating younger star-forming populations. This enhanced star formation in the nuclear disc qualitatively agrees with the observed bluer ends of the nuclear bar and nuclear spiral arms in the RGB image in Figure \ref{fig1}. Altogether, we have found that the nuclear structure of this galaxy displays properties similar to those of star-forming nuclear discs in the local Universe.

\section{Discussion}
\label{Sec:discussion}

We claim in this study that efficient bar-driven galaxy evolution has commenced by Cosmic Noon, resulting in the formation of a large nuclear disc. The size of the nuclear disc correlates with the length of the bar in nearby galaxies \citep[e.g.,][]{Seo_2019,Gadotti_2020}. As the bar drives gas inwards, the nuclear disc is believed to grow inside-out \citep[e.g.,][]{Kim_2012}, since the gas halts at successively larger radii, inducing bursts of star formation always more pronounced at the edge of the nuclear disc \citep[e.g.,][]{Bittner_2020}. In this study, we discover a large nuclear disc, $R_{\rm ND} \approx 1$ kpc and a long bar $\approx 5$ kpc, which sits above the nuclear disc size to bar length relation in \citet{Gadotti_2020}. \citet{Shlosman_1989} define the co-evolution of the nuclear disc and bar to be limited by the Inner Lindblad Resonance (ILR) at $0.1R_{bar}$. Here, we find a greater radius at $0.2R_{bar}$, suggesting that the nuclear disc has grown significantly and rapidly. Rapid substructure growth could be facilitated by the higher baryonic gas fractions observed at $z \approx 1.5$ \citep[see Fig. 8 in ][]{Zavadsky_2020}, which corresponds to $f_{\rm gas} \approx 0.4$ for the redshift and size of this galaxy. Gas is expected to play a significant role in the formation and evolution of barred galaxies, yet studies differ on the extent of its effects. Cosmological simulations find that, for strongly barred and unbarred galaxies, gas fractions are lower in the former \citep[][]{Rosas_Guevara_2020,Algorry_2017}, whilst \citet{Seo_2019} reported that cold stellar discs with large gas fractions formed bars earlier and more strongly, whereas high gas fractions in warm discs delayed bar formation. On the other hand, \citet{Athanassoula_2013b} shows that bar formation is more delayed in gas-rich discs than in gas-poor discs.

The recent study of \citet{Huang_2025} observed that extreme processes were at play in a massive barred galaxy ($4.5\times 10^{11} \rm M_\odot$) at $z = 2.467$. Through kinematic analysis, the authors found that the galaxy has a gas mass inflow rate of $579 \rm M_{\odot} yr^{-1}$. By comparison, this is more than an order of magnitude higher than the rates of 0.01 to $50 \rm M_{\odot}yr^{-1}$ found for barred galaxies in the local Universe \citep[][]{Hann_2009}. Hence, is the bar-driven evolution of CEERS-4031 as extreme as the processes reported in \citet{Huang_2025}? We make several assumptions to calculate the gas mass inflow rate of CEERS-4031, namely that the $ND/T$ luminosity ratio is equivalent to the $ND/T$ stellar mass ratio. Furthermore, we assume that the nuclear disc forms instantaneously after bar formation. Using the stellar mass and $ND/T$ ratio given in Table \ref{tab1}, we estimate the bar formation time given the gas inflow rate measured for local galaxies. For the most extreme and mean molecular gas inflow rates observed in \citet{Hann_2009}, 22.96 and $8.148 \rm M_{\odot}yr^{-1}$ (see their Table 9), we derive the formation times 0.42 and 1.17 Gyrs before the observation of this galaxy, corresponding to bar formation at the redshifts $z \approx 1.5$ to 1.7. Galaxies at Cosmic Noon are likely to be more gas-rich; thus, the more extreme gas inflow rates of locally barred galaxies would be more representative of this high-redshift galaxy. Alternatively, we calculate the bar formation time to be only $1.65\times 10^7 \rm yrs$ before the observation of this galaxy, assuming the inflow rate from \citet{Huang_2025}. This would require processes that rapidly form a nuclear disc, since some simulation studies find that nuclear discs form $\sim 10^8$ yrs after bar formation \citep[e.g.,][]{Athanassoula_1992,Emsellem_2015,Seo_2019}. Thus, it is unlikely that CEERS-4031 exhibits the same extreme dynamical processes as those of some other high-redshift galaxies. Altogether, these results suggest that the large gas fraction in CEERS-4031 has not influenced the formation and evolution of the bar, which is in line with the theoretical work of \citet{Fragkoudi_2025}.

A novel approach to age dating the bar of nearby disc galaxies is developed in \citet{deSaFreitas_2023,Freitas_2025}, where the authors predict that both bars and nuclear discs might be present since $z \sim 4$. The method assumes that the nuclear disc forms quickly after the bar, so differences in the star-formation histories of the nuclear and main discs reflect the bar's age. Until now, nuclear discs have been found only in nearby galaxies, and the assumption that nuclear discs form relatively quickly after the stellar bar has been tested in simulations. Our discovery of a nuclear disc in a high-redshift galaxy provides evidence that nuclear discs can form at early epochs, particularly the epoch of bar formation.

Nuclear structures larger than $\sim 600$ pc at $z \approx 1.6$ are resolvable with the NIRCam F200W filter, which is remarkably better than other space-based instruments \citep[see table C1 of ][]{Gadotti_2026}. CEERS-4031 has an extremely large nuclear disc with nuclear spiral arms and also appears to host a nuclear bar, all of which are resolved in NIRCam photometry. However, nuclear discs $< 0.8$ kpc in size would be missed. In a study of nearby nuclear discs, \citet{Gadotti_2025} reports that the distribution of $R_e$ ranges from 200 to 900 pc. In our study, the derived $R_e \approx 500$ pc agrees with the peak of the nearby galaxy distribution. As expected, the visually measured nuclear disc size is systematically larger than $R_e$. Our result is supported by the theoretical work of \citet{Kim_2012}, which finds nuclear spiral arms to be present in nuclear discs of type $x_2$ and $R_{\rm ND} > 0.6$ kpc. In the local Universe, up to $\approx50\%$ of the galaxies with stellar mass $\log_{10}(M_\star) > 10.5$\,M$_\odot$ (which corresponds to the galaxy in this study) host a nuclear bar \citep[e.g.,][]{Erwin_2024}. Population statistics will improve our understanding of the environments which promote bar-driven evolution and unveil the role of nuclear discs in AGN fuelling during the peak of cosmic AGN activity. However, such studies will only reflect the properties of the largest nuclear discs due to the aforementioned resolution effects. The MICADO instrument at the Extremely Large Telescope will be able to resolve most, if not all, nuclear discs at all redshifts \citep{Gadotti_2026}.

Our discovery of a nuclear disc at Cosmic Noon provides evidence that we need to reevaluate the terminology of `secular’ in the context of bar-driven evolution. Whilst some aspects of bar-driven evolution may be secular, other processes within the bar radius may be considered to be rapid, especially regarding gas inflow and the building of nuclear structures. Furthermore, we measure the nuclear disc size to be on the larger end of the size distribution in nearby galaxies, indicating that gas has been efficiently transported and accumulated in the central region.

\section{Conclusions}
\label{Sec:conclusions}
We discover the most distant nuclear disc within a barred galaxy to date, at $z = 1.461$, implying unequivocal evidence for bar-driven galaxy evolution during early epochs. The nuclear disc has a radius of $R_{{\rm ND}} \approx 1$ kpc and shows substantial evidence of hosting a nuclear bar and nuclear spiral arms, and exhibits a near-exponential radial profile, thus consistent with nuclear discs in nearby galaxies. Within the nuclear disc region, we observe heightened star formation and younger stellar populations, indicating bar-driven gas inflow. This nuclear disc thus exhibits hallmark signatures of bar-driven evolution in nearby galaxies and provides evidence for the rapid maturation of galaxy discs and bars at Cosmic Noon.

\section*{Acknowledgements}

We thank the referee for their thoughtful comments. This work was supported by STFC grants ST/X508354/1, ST/T000244/1 and ST/X001075/1. This work used the DiRAC@Durham facility managed by the Institute for Computational Cosmology on behalf of the STFC DiRAC HPC Facility (www.dirac.ac.uk). The equipment was funded by BEIS capital funding via STFC capital grants ST/K00042X/1, ST/P002293/1, ST/R002371/1 and ST/S002502/1, Durham University, and STFC operations grant ST/R000832/1. DiRAC is part of the National e-Infrastructure. TK is supported by Basic Science Research Program through the National Research Foundation of Korea (NRF) funded by the Ministry of Education (RS-2025-25399934).

\section*{Data Availability}
 
The specific observations analysed can be accessed via \url{https://doi.org/10.17909/xm8m-tt59}. This work used \textsc{Astropy} \cite[][]{Astropy_2013}{}{}, \textsc{SExtractor} \cite[][]{Bertin_1996}, \textsc{PHOTUTILS} \cite[][]{Bradley_2022}{}{}, \textsc{imfit} \cite[][]{Erwin_2015} and \textsc{expanse} \citep[e.g.,][]{harvey2025}.



\bibliographystyle{mnras}
\bibliography{LeConte} 








\bsp	
\label{lastpage}
\end{document}